\documentclass{elsart}

\usepackage{graphicx}
\usepackage{amssymb}

\begin{document}

\begin{frontmatter}

\title{Octahedral Tilting in $ {\rm \bf ACu_3Ru_4O_{12}} $
       (A=Na,Ca,Sr,La,Nd)}

\author{U.~Schwingenschl\"ogl\corauthref{cor1}},
\ead{Udo.Schwingenschloegl@physik.uni-augsburg.de}
\author{V.~Eyert},
\author{U.~Eckern}
\corauth[cor1]{Corresponding author. Fax: 49-821-598-3262}
\address{Institut f\"ur Physik, Universit\"at Augsburg, 86135 Augsburg,
         Germany}

\begin{abstract}
The perovskite-like compounds $ {\rm ACu_3Ru_4O_{12}} $ (A=Na,Ca,Sr,La,Nd) 
are studied by means of density functional theory based electronic structure 
calculations using the augmented spherical wave (ASW) method. 
The electronic properties are strongly influenced by covalent type bonding 
between transition metal $ d $ and oxygen $ p $ states. The characteristic 
tilting of the $ {\rm RuO_6} $ octahedra arises mainly from the 
Cu--O bonding, allowing for optimal bond lengths between these two atoms. 
Our results provide a deeper understanding of octahedral tilting as a universal
mechanism, applicable to a large variety of multinary compounds. 
\end{abstract}

\end{frontmatter}

\section{Introduction}

Despite its striking simplicity the perovskite structure, $ {\rm ABX_3} $,  
contains numerous crystallographic variations giving room for a huge 
class of compounds \cite{woodward97,woodward97b}. Among these the vast majority 
comprises oxides and fluorides \cite{goodenough70}, but also chlorides, 
hydrides, oxynitrides, and sulfides. The great interest in these materials 
is motivated by a variety of exciting dielectric, magnetic, electrical, 
optical, and catalytic properties, with several technological applications.\\  
For optimal tailoring of materials much work on the perovskite-related 
compounds concentrates on the interrelations between deviations from the 
ideal perovskite structure and physical properties. These crystallographic 
deviations can be grouped into three different mechanisms 
\cite{woodward97,woodward97b}: While 
distortions of the characteristic $ {\rm BX_6} $ octahedra and cation 
displacements within the octahedra are mainly driven by electronic 
instabilities of the octahedral metal ion, octahedral tiltings represent 
the most common deviation and are observed when the A cation is too 
small for filling the space between regularly ordered octahedra.  
In this situation, octahedral tilting represents the lowest energy 
distortion, since it allows to adjust the A--O distances while 
leaving the first coordination sphere of the M cation intact, and 
changing mainly the soft M--O--M bond angle rather than the M--O 
bond length.\\ 
Of particular interest in this context are the perovskite-like compounds  
of the kind $ {\rm AA'_3B_4O_{12}} $, where the B--O--B bond angle is 
the relevant quantity in order to optimize both the $ {\rm A} $--O and
$ {\rm A'} $--O bond 
lengths. Studying these materials thus promises deeper insight into the 
mechanisms driving the octahedral tilting. In addition, this material 
class shows extraordinary physical properties. 
While $ {\rm CaCu_3Mn_4O_{12}} $ is a ferromagnetic semiconductor with high 
Curie temperature and large magnetoresistance \cite{zeng99},
$ {\rm CaCu_3Ti_4O_{12}} $ is known for its 
unusually high low-frequency dielectric constants 
\cite{subramanian00,homes01,subramanian02}. In contrast, the metallic 
ruthenates of the type $ {\rm ACu_3Ru_4O_{12}} $ (A=Na,Ca,La) are Pauli 
paramagnets \cite{labeau80} and
display valence degeneracy \cite{subramanian02}.
The crystal structures of the latter class have been studied recently by
EXAFS and XRD, revealing considerable deviations of 
the measured interatomic distances from those expected from a bond 
valence approach \cite{ebbinghaus02}.\\
The present study investigates the relationship between the 
octahedral tiltings and the electronic properties of the ruthenates 
$ {\rm ACu_3Ru_4O_{12}} $ (A=Na,Ca,Sr,La,Nd). The discussion is based 
on electronic structure calculations within density functional 
theory, which are known for their predictive power (see e.g.\ 
Refs.\ \cite{bookmat,bookdft} and references therein). As a result, 
the specific bond lengths as well as the discrepancies between the 
experimental interatomic distances of the class $ {\rm ACu_3Ru_4O_{12}} $
and those proposed by the bond valance approach can be understood from the
electronic properties and the type of chemical bonding.

\section{Crystal Structure}

\begin{figure}
\centering 
\includegraphics*[width=0.7\textwidth]{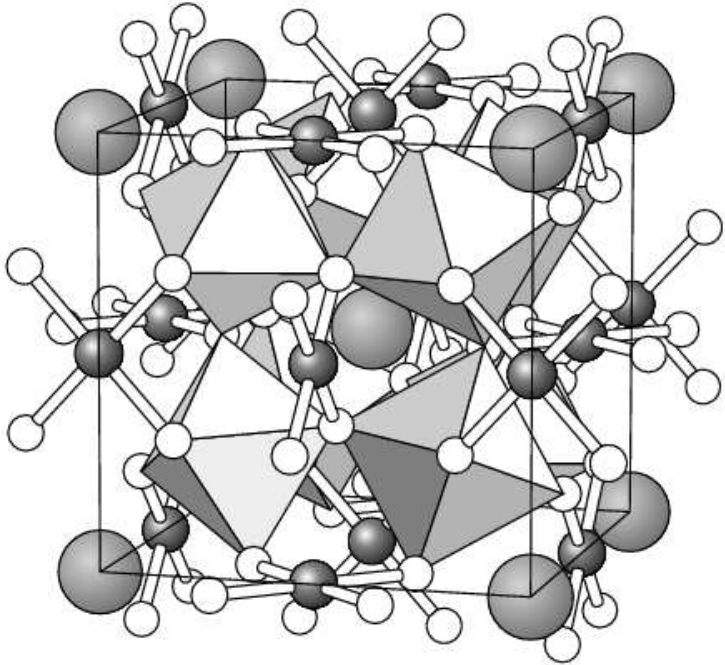}
\caption{Crystal structure of $ {\rm AA'_3B_4O_{12}} $. Large and medium 
         size spheres denote A- and Cu-atoms, respectively, while octahedra 
         represent the $ {\rm BO_6} $ units.}
\label{fig:cryst1}
\end{figure}

The compounds of the class $ {\rm AA'_3B_4O_{12}} $ crystallize in a 
body-centered cubic $ 2 \times 2 \times 2 $ superstructure of the 
simple cubic perovskite structure with space group $ Im\bar{3} $ 
(No.\ 204). The atoms are located at the following Wyckoff positions:  
A ($ 2a $) at (0,0,0), $ {\rm A'} $ ($ 6b $) at $ (\frac{1}{2},0,0) $, 
B ($ 8c $) at $ (\frac{1}{4},\frac{1}{4},\frac{1}{4}) $, and O ($ 24g $) 
at $ (x,y,0) $.  The crystal structure is shown in Fig.\ \ref{fig:cryst1}, 
where the $ {\rm BO_6} $ octahedra are easily identified. The tilting of 
the octahedra is controlled by the oxygen position, which, for all the 
ruthenates under study, is specified by $ x_{\rm O} \approx 0.175 $ and 
$ y_{\rm O} \approx 0.305 $. The deviation from the ideal values  
$ x_{\rm O} = y_{\rm O} = 0.25 $ causes a rotation of the octahedra around 
the [111] axis. As a consequence, the A sites are partitioned
into the A site at (0,0,0) and the $ {\rm A'} $ site, which takes three 
quarters of the original A positions. In particular, the $ {\rm A'} $ 
sites are strongly affected by the oxygen atom shifts away from the ideal 
position and thus their surrounding experiences the largest distortion. 
As a consequence, three different $ {\rm A'} $--O distances of $ \approx $ 
2.0\,\AA, 2.8\,\AA, and 3.2\,\AA\ appear, each forming an approximately 
square-planar coordination and thus providing an ideal geometry for a 
Jahn-Teller active ion as e.g.\ $ {\rm Cu^{2+}} $ \cite{ebbinghaus02}. 
The $ {\rm CuO_4} $ squares can be easily identified in Fig.\ 
\ref{fig:cryst1}.\\
For the ruthenates $ {\rm ACu_3Ru_4O_{12}} $ (A=Na,Ca,Sr,La,Nd) studied 
in the present work, crystallographic data were determined by Ebbinghaus 
{\em et al.}\ \cite{ebbinghaus02}. The lattice constants are similar to 
those reported by Labeau {\em et al.}\ \cite{labeau80} as well as
Subramanian and Sleight \cite{subramanian02}. Both the lattice constants and
oxygen parameters of these compounds show only slight deviations and amount to 
$ a \approx 7.4 $\,\AA\ as well as $ x_{\rm O} \approx 0.175 $ and 
$ y_{\rm O} \approx 0.305 $ as mentioned above.

\section{Method of Calculation}

The electronic structure calculations were performed within the framework 
of density functional theory and the local density approximation using the 
augmented spherical wave (ASW) method \cite{williams79,eyert00}. In order 
to represent the correct shape of the crystal potential in the large 
voids, additional augmentation spheres were inserted into the open crystal 
structures. Optimal augmentation sphere positions as well as radii of all 
spheres were found automatically by the sphere geometry
optimization (SGO) algorithm \cite{eyert98}.
The basis sets comprised Cu $4s$, $4p$, $3d$, Ru $5s$, $5p$, $4d$, $4f$, and
O $2s$, $2p$ as well as states of the additional augmentation spheres. In 
addition, Na $3s$, $3p$, $3d$; Ca  $4s$, $4p$, $3d$, $4f$; Sr $5s$, $5p$, 
$4d$, $4f$; La/Nd $6s$, $6p$, $5d$, $4f$ states were used for the A cation. 
The Brillouin zone sampling was done using an increasing number of ${\bf k}$ 
points within the irreducible wedge ranging from 17 to 1255 points to ensure 
convergence of the results with respect to the $ {\bf k} $-space grid. 
In addition to the analysis of the (partial) densities of states we will also 
address the question of chemical bonding in terms of the covalent bond energy, 
which was recently proposed \cite{boernsen99} as an extension to the concept of
the crystal orbital overlap population (COOP) \cite{hoffmann88}. In short, 
the covalent bond energy covers those contributions to the total energy of 
a crystal, which arise from the hybridization of orbitals located at different 
atoms.

\section{Results and Discussion}

\begin{figure}
\centering 
\includegraphics[width=0.98\textwidth]{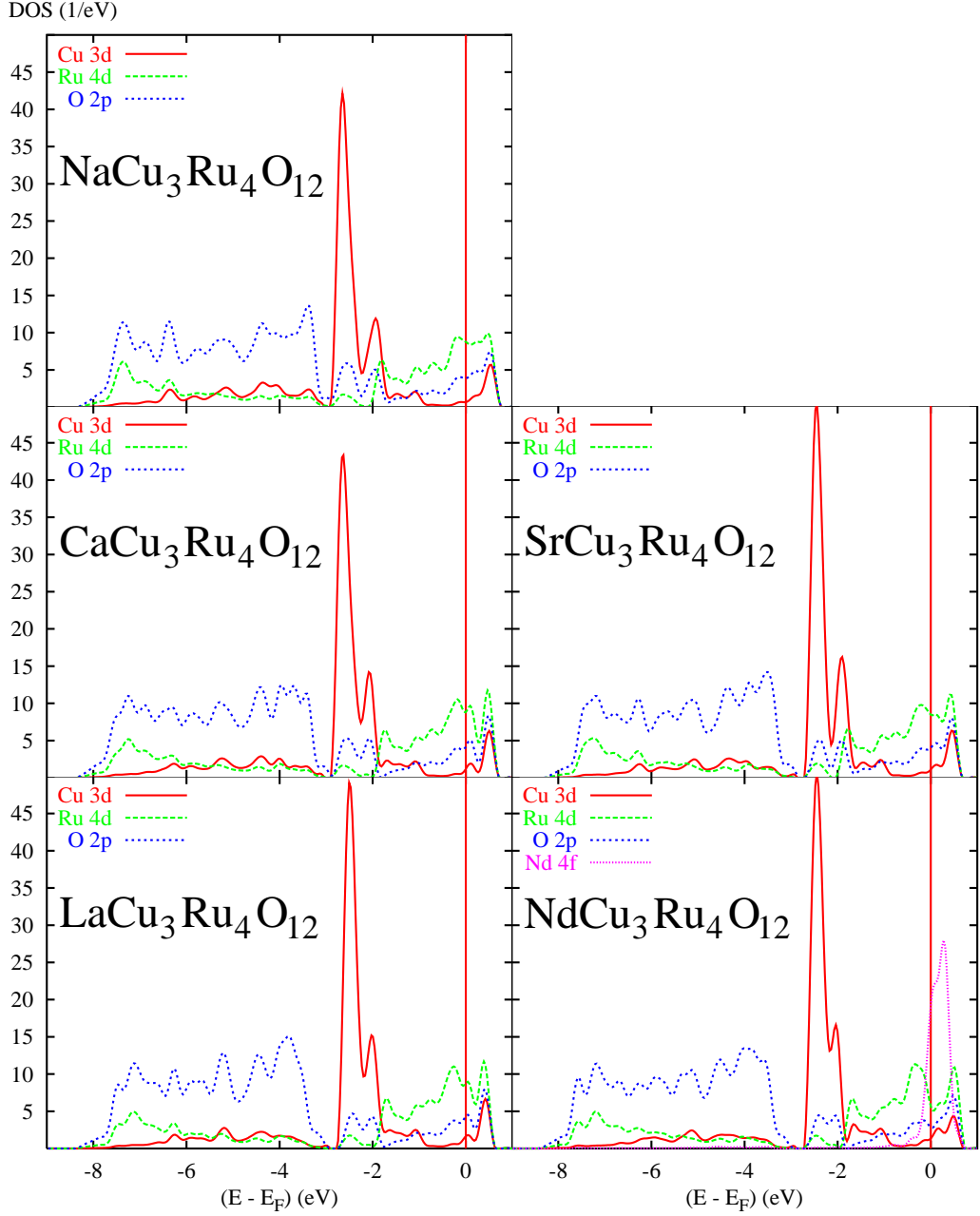}
\caption{Partial Cu $ 3d $, Ru $ 4d $, and O $ 2p $ densities of states 
         (DOS) of $ {\rm ACu_3Ru_4O_{12}} $ per formula unit. $ 4f $ 
         states have been included for the Nd-compound.} 
\label{fig:res1}
\end{figure}

\begin{figure}
\centering
\includegraphics[width=0.60\textwidth,clip]{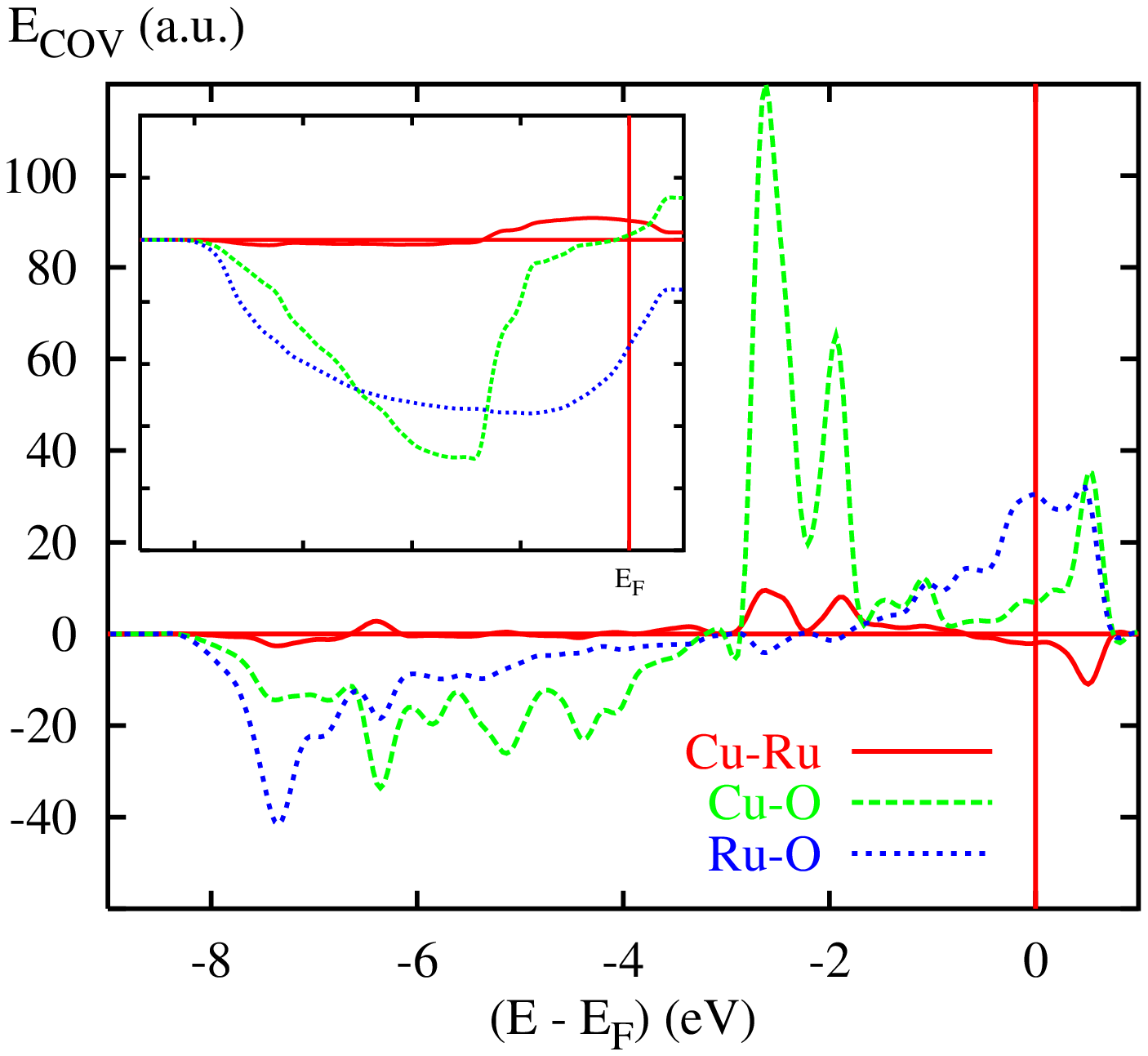}
\caption{Covalent bond energy curves for $ {\rm  NaCu_3Ru_4O_{12}} $. 
         The corresponding integrated covalent bond energies are given  
         in the inset.}
\label{fig:res2}
\end{figure}

\begin{figure}
\centering
\includegraphics[width=0.98\textwidth,clip]{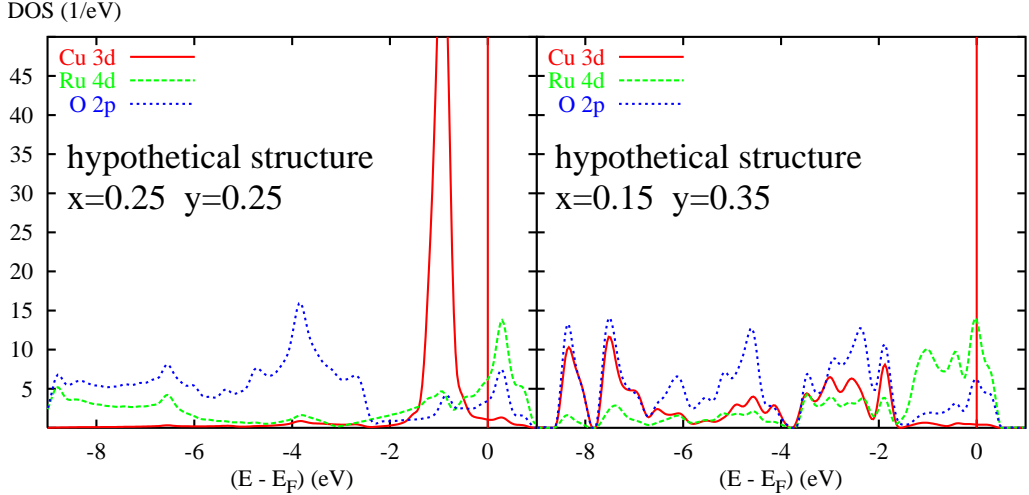}
\caption{Partial Cu $ 3d $, Ru $ 4d $, and O $ 2p $ densities of states 
         (DOS) of hypothetically distorted variants of 
         $ {\rm NaCu_3Ru_4O_{12}} $ per formula unit.} 
\label{fig:res3}
\end{figure}

Partial densities of states (DOS) for all five compounds under consideration 
are displayed in Fig.\ \ref{fig:res1}. 
The electronic structure in the energy window shown is dominated by 
Cu $ 3d $, Ru $ 4d $, and O $ 2p $ states. In addition, Nd $ 4f $ states 
cause a sharp peak at and slightly above $ {\rm E_F} $ for  
$ {\rm NdCu_3Ru_4O_{12}} $. The gross features are very similar for 
all compounds. Three energy ranges can be distinguished. In the broad 
interval from about -8 to -3\,eV oxygen $ 2p $ states dominate but are 
complemented by considerable contributions from the transition metal 
$ d $ states due to covalent bonding. Whereas in the energy range from 
-3 to -2\,eV a sharp Cu $ 3d $ peak appears, electronic states between 
-2 and 0.5\,eV, and hence metallic conductivity, derive mainly from 
broader Ru $ 4d $ bands. Covalent type bonding leads to finite oxygen 
$ 2p $ contributions above -3\,eV. This is confirmed by covalent bond 
energy curves calculated for the sodium compound as shown in Fig.\ 
\ref{fig:res2}. 
Both the $ {\rm Cu} $--$ {\rm O} $ and the $ {\rm Ru} $--$ {\rm O} $ 
curves are negative and positive at energies below and above -3\,eV,
respectively, indicative of bonding and antibonding states in the 
energy regions, where the O $ 2p $ and the transition metal $ d $ 
states dominate. However, due to the higher $ d $ electron count of Cu 
as compared to Ru, the $ {\rm Cu} $--$ {\rm O} $ antibonding states are 
filled, leading to a vanishing contribution of these bonds to the chemical 
stability. In contrast, the integrated $ {\rm Ru} $--$ {\rm O} $ covalent bond 
energy curve at $ {\rm E_F} $ has a finite value indicative of the 
stabilizing net contribution of this bond. Finally, metal-metal bonding 
results in small peaks between -3 and -2\,eV and above the Fermi energy.\\
All states which are not displayed in Fig.\ \ref{fig:res1} play only a 
very small role in the energy interval shown. In particular, apart from 
the Nd $ 4f $ states, the A cation electrons do not contribute to the 
covalent-type bonding but are distributed over the solid. As a consequence, 
changes at the A cation site show up mainly in the modified total electron 
count.\\
Ebbinghaus {\em et al.} \cite{ebbinghaus02} related the measured atomic 
distances for $ {\rm ACu_3Ru_4O_{12}} $ (A=Na,Ca,Sr,La,Nd) to the 
predictions of the bond valence model. The latter is an empirical model, 
which connects the valences and the mutal distances of the atoms 
\cite{brown85}. It starts out from ionic configurations but looses accuracy 
in covalently bonded crystals. For the divalent A cations the Ru-O distances 
were found to agree with the expected values. In the case of A=Na the bond 
length is longer, while for A=La,Nd it is shorter than predicted by the bond 
valence model.\\
In order to understand the influence of octahedral tilting on the electronic 
properties and to analyze the mentioned discrepancies reported by Ebbinghaus
{\em et al.} we performed additional calculations using hypothetical 
crystal structures. In these setups the oxygen atoms were shifted away 
from the measured positions to the ideal position $ x_{\rm O} = 0.25 $, 
$ y_{\rm O} = 0.25 $ and to $ x_{\rm O} = 0.15 $, $ y_{\rm O} =0.35 $. 
These positions correspond to zero tilting of the octahedra and to an 
exaggerated tilting, respectively. Note that shifting the oxygen position 
mainly affects the Cu--O bond lengths, while the effect for the Ru--O and 
A--O distances is of second order. The partial DOS for the 
hypothetical structures of $ {\rm NaCu_3Ru_4O_{12}} $ are displayed in 
Fig.\ \ref{fig:res3}. 
Considerable changes as compared to the results for the experimental crystal 
structure (see Fig.\ \ref{fig:res1}) are observed. Turning to the ideal 
structure first, we observe (i) a considerable broadening of the O $ 2p $ 
dominated states and (ii) a striking sharpening of the Cu $ 3d $ states. 
Both effects are readily understood from the oxygen shifts. The latter cause 
an increase of the Cu--O distances and, to a lesser degree, a decrease of the 
Ru--O distances, leading to a slightly larger $ 2p $--$ 4d $ overlap. As a 
consequence, larger Ru $ 4d $ 
contributions at energies below -3\,eV are observed, whereas the Cu $ 3d $ 
fraction is almost negligible in this region. At the same time, Ru $ 4d $ 
and O $ 2p $ admixture to the sharp Cu $ 3d $ peak is likewise very small. 
To sum up, shifting the oxygen atoms to their ideal positions leads to an 
effective decoupling of the compound into $ {\rm RuO_6} $ octahedra and 
Cu atoms, which are effectively separated and thus do no longer form a 
stable solid.\\
On the contrary, for the hypothetical structure with the oxygen shift away 
from the ideal position exaggerated (i) the oxygen states have moved to 
lower energies and (ii) the Cu $ 3d $ states do no longer form a distinct 
peak but have attained a rather itinerant character. The latter goes along 
with an increased similarity of the Cu $ 3d $ and O $ 2p $ partial DOS. 
The increased covalent type bonding between these two states is also reflected 
by large contributions to the covalent bond energy (not shown), negative below
-4\,eV and positive above, leading to a net antibonding integral at 
$ {\rm E_F} $.  The increased $ 2p $--$ 3d $ bonding is in contrast to a slight 
decrease of the bonding between the Ru $ 4d $ and O $ 2p $ states due to 
the larger extension of the $ {\rm RuO_6} $ octahedra. Thus the 
hypothetical structure with exaggerated oxygen shifts likewise suffers 
from an increase of antibonding at the expense of bonding states.\\
Together with the instability of the hypothetical structure with 
ideal oxygen positions we conclude that the experimental 
structure is due to an optimal balance of different types of bonds, namely 
$ 2p $--$ 3d $ and $ 2p $--$ 4d $. This balance is less affected by the A 
cations, which do not take part in covalent bonding but have lost their 
outer electrons. For this reason, from the point of view of chemical 
bonding, there is no means to optimize the A--O bond length. Moreover, as 
for the Ru--O distance, the A--O distance is only to a small degree 
affected by the tilting. Finally, the different types of bonding present 
in these compounds give an additional explanation for the deviations of 
the experimental interatomic distances from those expected from the bond 
valence model.

\section{Conclusions}

The electronic properties of the perovskite-like ruthenates 
$ {\rm ACu_3Ru_4O_{12}} $ are governed by strong covalent bonding between 
the transition metal $ d $ and the oxygen $ 2p $ electrons. While 
Ru--O bonds via the corresponding bond-lengths influence mainly the 
size of the octahedra and hence the lattice constant, the octahedral 
tilting is to a large part due to the Cu--O bonding.\\
As compared to the ideal perovskite structure the tilting of the 
characteristic $ {\rm RuO_6} $ octahedra has only little effect on 
the Ru--O bonding. In contrast, the Cu--O distances undergo serious 
changes, leading eventually to the square-planar coordination of copper. 
Finally, the unusual A--O distances and the failure of the bond valence 
approach are due to the fact that their response to the tilting is only 
small and the structure does not offer any possibility to simultaneously 
optimize the A--O and the Cu--O bond lengths. \\
In conclusion, octahedral tilting as discussed for $ {\rm ACu_3Ru_4O_{12}} $,
arises from the interplay of covalent bonding between different atomic 
species, leading to optimized bond lengths between all atoms involved. This 
general principle is of rather universal nature and applies to a large 
variety of compounds.

\section*{Acknowledgements}
Fruitful discussions with S.\ G.\ Ebbinghaus and A.\ Weidenkaff are 
gratefully acknowledged. This work was supported by the Deutsche 
Forschungsgemeinschaft through SFB 484.

\end{document}